# On Inference of Weitzman Overlapping Coefficient in Two Weibull Distributions

**Omar M. Eidous** and **Hala Y. Maqableh**

omarm@yu.edu.jo            halakhaled1711@gmail.com

**Department of statistics - Yarmouk University**

**Irbid - Jordan**

**2024**

## Abstract

Studying overlapping coefficients has recently become of great benefit, especially after its use in goodness-of-fit tests. These coefficients are defined as the amount of similarity between two statistical distributions. This research examines the estimation of one of these overlapping coefficients, which is the Weitzman coefficient $\Delta$, assuming two Weibull distributions and without using any restrictions on the parameters of these distributions. We studied the relative bias and relative mean square error of the resulting estimator by implementing a simulation study. The results show the importance of the resulting estimator.

# 1. Introduction

Overlapping coefficients are used to determine similarity between two populations. If we want to make comparative inferences about two populations, we will look at what are called measures of similarity or measures of dissimilarity. In fact, when we discuss this type of measure, we should pay attention to the overlapping (OVL) coefficients. Therefore, OVL coefficients are measure of agreement or similarity between two probability statistical distributions.

There are five main measures of OVL, which are: Matusita ($\rho$), Morisita ($\lambda$), Pianka (PI), Kullback –Leibler (KL) and Weitzman ($\Delta$) coefficients. They are defined as follows:

Assuming $f_1(x)$ and $f_2(x)$ are continuous probability density functions, the five OVL coefficients are:

1. Matusita coefficient ( 1955 ):
$$\rho = \int \sqrt{f_1(x)f_2(x)}\, dx$$

2. Morsita coefficient ( 1959 ):
$$\lambda = \frac{2\int f_1(x)f_2(x)\, dx}{\int [f_1(x)]^2\, dx + [f_2(x)]^2\, dx}$$

3. Pianka coefficient (Chaubey *et al.*, 2008):
$$PI = \frac{\int f_1(x)f_2(x)\, dx}{\sqrt{\int [f_1(x)]^2 dx \int [f_2(x)]^2 dx}}$$

4. Kullback-Leibler coefficient (Dhaker *et al.*, 2019)
$$KL = \frac{1}{1 + \int (f_1(x) - f_2(x))\log(f_1(x)/f_2(x))\, dx}$$

5. Weitzman coefficient ( 1970 ):
$$\Delta = \int min\{f_1(x), f_2(x)\}dx$$

The Weitzman coefficient $\Delta$ is a widely used and more clearly defined than the other coefficients, which represents the area of intersection between two probability density functions (Inman and Bradley, 1989). Our interest in this paper is only the Weitzman coefficient $\Delta$. The main objective is to estimate $\Delta$ assuming two Weibull distributions and without using any restrictions on the parameters of these distributions. If the value of any of the above five coefficients is 1 (i.e. $OVL = 1$) then $f_1(x) = f_2(x)$. If $OVL = 0$, then the supports of the two densities $f_1(x)$ and $f_2(x)$ have no interior points.

Overlap measures are applied in different areas like, genetic (Federer et al. 1963), ecology (Pianka, 1973), income (Gastwirth, 1975), reliability analysis (Ichikawa, 1993 and Dhaker *et al.* 2019) and goodness of fit test (Alodat *et al.*, 2022).

## 2. Weibull Distribution and OVL Coefficients

The Weibull distribution is a continuous probability density function. This distribution attracted the interest of statisticians due to its advantages such as its flexibility to model data sets in many fields of applied statistics, like, lifetime data, economics and business administration data, engineering studies data and wind power data (see Rinne, 2008 and Wais, 2017). Let $X$ be a continuous random variable that follows a Weibull distribution with a scale parameter $\alpha$ and a shape parameter $\beta$ then the $pdf$ of $X$ is,

$$f(x) = \frac{\beta}{\alpha}\left(\frac{x}{\alpha}\right)^{\beta-1} e^{-(x/\alpha)^\beta}, \qquad x > 0, \qquad \alpha, \beta > 0.$$

This will be denoted by $X \sim We(\alpha, \beta)$. Some well-known statistical distributions are special cases of $We(\alpha, \beta)$, including the exponential distribution, which is obtained if $\beta = 1$, and the Rayleigh distribution, which is obtained if $\beta = 2$ and $\alpha = \sqrt{2}\sigma$ (Rinne, 2008).

Let $X \sim W(\alpha_1, \beta_1)$ and $Y \sim W(\alpha_2, \beta_2)$ where $X$ and $Y$ are independent random variables. Assume that $\beta_1 = \beta_2 = 1$, the OVL coefficients $\lambda$, $\rho$ and $\Delta$ were studied by Madhuri *et al.* (2001) and Samawi and Al-Saleh (2008), who also studied the effect of sampling plan on these OVL coefficients. Under the assumption, $\beta_1 = \beta_2 = \beta$, the coefficients $\lambda$, $\rho$ and $\Delta$ were studied by Al-Saidy *et al.* (2005), while the coefficients $PI$ and $KL$ were studied by Eidous and Abu Al-Hayja'a (2023c). Finally, without using any assumptions on the parameters of Weibull distributions, Eidous and Abu Al-Hayja'a (2023a) were concerned with the coefficients $\lambda$ and $\rho$, while the study by Eidous and Abu Al-Hayja'a (2023b) focused on the coefficient $\Delta$. In the last both studies, the numerical integration approximation method was used to study the various coefficients.

There are researches that have studied OVL coefficients under statistical distributions other than Weibull distributions. Inman and Bradley (1989), Mulekar and Mishra (1994), Eidous and Al-Daradkeh (2022) and Eidous and Al-Shourman (2023) considered the case of normal distributions.

Helu and Samawi (2011) investigated the OVL coefficients of Lomax distributions with different sampling procedures. Parametric methods for estimating the confidence interval for $\Delta$ have been studied by Wang and Tiana (2017) who also proposed methods for estimating the confidence interval for $\Delta$ undera variety of distributions, including the normal distribution.

There are also some nonparametric studies that were concerned with studying OVL coefficients, which can be found in the literature. These studies do not assume any specific statistical distributions for the phenomenon under study. See for example, Schmid and Schmidt (2005), Pastore (2018), Eidous and AL-Talafha (2022), Eidous and Ananbeh (2024a) and Eidous and Ananbeh (2024b).

## 3. Main Results

Let $X_1, X_2, \ldots, X_{n_1}$ be a random sample from $W(\alpha_1, \beta_1)$, and let $Y_1, Y_2, \ldots, Y_{n_2}$ be another random sample from $W(\alpha_2, \beta_2)$, where the two samples are independent. If $\beta_1 = \beta_2 = \beta$, let $\hat{\alpha}_1, \hat{\alpha}_2$, and $\hat{\beta}$ are the maximum likelihood estimators (MLEs) of $\alpha_1, \alpha_2, \beta$ respectively. If there is no restriction about the distributions parameters, let $\hat{\alpha}_1, \hat{\alpha}_2, \hat{\beta}_1$ and $\hat{\beta}_2$ are the MLEs of $\alpha_1, \alpha_2, \beta_1$ and $\beta_2$ (See Eidous and Abu Al-Hayja'a, 2023a).

To estimate the Weitzman Coefficient $\Delta$ based on these two random samples, we first express the coefficient $\Delta$ as follows:

To simplify the notations, let $f_1(X) = W(\alpha_1, \beta_1)$ and $f_2(X) = W(\alpha_2, \beta_2)$. Consider $min\{f_1(X), f_2(X)\}/f_1(X)$ as a function of $X$ and $min\{f_1(Y), f_2(Y)\}/f_2(Y)$ as a function of $Y$. Now,

$$E\left(\frac{min\{f_1(X), f_2(X)\}}{f_1(X)}\right) = \int_0^\infty \frac{min\{f_1(x), f_2(x)\}}{f_1(x)} f_1(x)\, dx$$

$$= \int_0^\infty min\{f_1(x), f_2(x)\}\, dx$$

$$= \Delta$$

and

$$E\left(\frac{min\{f_1(Y), f_2(Y)\}}{f_2(Y)}\right) = \int_0^\infty \frac{min\{f_1(y), f_2(y)\}}{f_2(y)} f_2(y)\, dy$$

$$= \int_0^\infty min\{f_1(y), f_2(y)\}\, dy$$

$$= \int_0^\infty min\{f_1(x), f_2(x)\}\, dx$$

$$= \Delta.$$

Also, $\Delta$ can be expressed as follows,

$$\Delta = \frac{1}{2}\left[E\left(\frac{min\{f_1(X), f_2(X)\}}{f_1(X)}\right) + E\left(\frac{min\{f_1(Y), f_2(Y)\}}{f_2(Y)}\right)\right]$$

If $\hat{f}_1(X) = W(\hat{\alpha}_1, \hat{\beta}_1)$ and $\hat{f}_2(X) = W(\hat{\alpha}_2, \hat{\beta}_2)$ then $\Delta = E\left(min\{f_X(X), f_Y(X)\}/f_X(X)\right)$ can be estimated by using the method of moments as given below,

$$\hat{\Delta} = \frac{1}{n_1}\sum_{k=1}^{n_1}\left(\frac{min\{\hat{f}_1(X_k), \hat{f}_2(X_k)\}}{\hat{f}_1(X_k)}\right)$$

Also, $\Delta = E(min\{f_X(Y), f_Y(Y)\}/f_Y(Y))$ can be estimated by,

$$\hat{\Delta} = \frac{1}{n_2}\sum_{k=1}^{n_2}\left(\frac{min\{\hat{f}_1(Y_k), \hat{f}_2(Y_k)\}}{\hat{f}_2(Y_k)}\right).$$

The average of the last two estimators can be considered as the third estimator for $\Delta$, which is given by,

$$\hat{\Delta} = \frac{1}{2n_1}\sum_{k=1}^{n_1}\left(\frac{min\{\hat{f}_1(X_k), \hat{f}_2(X_k)\}}{\hat{f}_1(X_k)}\right) + \frac{1}{2n_2}\sum_{k=1}^{n_2}\left(\frac{min\{\hat{f}_1(Y_k), \hat{f}_2(Y_k)\}}{\hat{f}_2(Y_k)}\right).$$

Note that if we assume that $\beta_1 = \beta_2 = \beta$ then $\hat{f}_1(X) = W(\hat{\alpha}_1, \hat{\beta})$ and $\hat{f}_2(X) = W(\hat{\alpha}_2, \hat{\beta})$.

The performances of these three estimators are investigated in a preliminary simulation study. The results show that the performance of $\hat{\Delta}$ (last version) is more stable than the first two versions.

## 4. Simulations

A simulation study is conducted to compare the performances of the proposed estimator $\hat{\Delta}$ (last version in the previous section) of $\Delta$ with some existing counterparts that developed in the literature. In particular, the nonparametric kernel estimator is considered for this purpose, which is denoted by $\hat{\Delta}_k$. It is important to note that the kernel estimator is a general estimator, which does not require any assumptions about the shape of the underlying sample distribution (see, Eidous and Al-Talafha, 2022).

To cover most possible cases in practical applications, the two independent samples $x_1, x_2, \ldots, x_{n_1}$ and $y_1, y_2, \ldots, y_{n_2}$ are simulated from 12 pairs of Weibull distributions. From these pairs, four pairs with the same scale parameters (i.e. $\alpha_1 = \alpha_2$), four pairs with the same shape parameters (i.e. $\beta_1 = \beta_2$) and four pairs with different scale parameters and different shape parameters were selected. Although these choices seem arbitrary, the goal was to allow the exact overlapping coefficient values to vary from small (near 0) to large (near 1). In Tables (1)-(3), the parameter values are shown along with the exact value of $\Delta$ for each pair. To study the effect of sample sizes on the behavior of each estimator, for following ample sizes were taken $(n_1, n_2) = (10,10), (20,30), (30,30), (50,50), (100,200)$.

Numerical results were calculated based on a thousand iterations ($R = 1000$) using Mathematica, Version 7. For each estimator, we calculated the Relative Bias (RB), Relative Root Mean Square Error (RRMSE) and Efficiency (EFF), which can be defined as follows,

$$RB = \frac{\hat{E}(estimator) - exact}{exact},$$

$$RRMSE = \frac{\sqrt{\widehat{MSE}(estimator)}}{exact}$$

and

$$EFR = \frac{\widehat{MSE}(kernel)}{\widehat{MSE}(proposed\ estimator)}.$$

Note that if $\widehat{\Delta}$ is the estimator of $\Delta$ and if $\widehat{\Delta}_{(j)}$ is the value of $\widehat{\Delta}$ computed based on a sample of iteration $j, j = 1, 2, …, R = 1000$ then,

$$\widehat{E}(\widehat{\Delta}) = \sum_{j=1}^{R} \widehat{\Delta}_{(j)}/R$$

and

$$\widehat{MSE}(\widehat{\Delta}) = \sum_{j=1}^{R}\left(\widehat{\Delta}_{(j)} - E(\widehat{\Delta})\right)^2 /R.$$

Note also that the kernel estimator requires specifying two quantities, the first is the kernel function, which took the standard normal function, and the other quantity is the smoothing parameter, which is calculated using the rule,

$$1.06\ S\ n^{-1/5}$$

where $S$ is the usual standard deviation for the interested sample. It is worth noting here that there are other ways to calculate the smoothing parameter (see Eidous *et al.*, 2010). However, we found that the performance of the kernel estimator using the above rule is very acceptable.

**5. Simulation Results**

All computations and outputs of the simulation study are presented in Tables (1-3). From these simulation results, we can conclude the following:

1. It is obvious that $|RB|$s that are associated with the kernel estimator $\widehat{\Delta}_k$ are large compared with other the proposed estimator, especially for small samples sizes. Most $RBs$ values of the kernel estimates are negative, which indicates that -on the average- the kernel estimates underestimate the true value of the corresponding coefficient. It appears that the problem of underestimate is a problem associated with the kernel estimates, even in other fields such as line transect method (Eidous, 2009, 2011 and 2012) and degradation methods (Ba Dakhn *et al.*, 2017 and Arif and Eidous, 2017).
2. The values $|RB|$s of the different proposed estimate are much smaller than that of the kernel estimate for almost all considered cases.
3. As the samples sizes increase the RRMSE of the two estimators decrease. This is a good sign for the consistency of the estimators that considered in this study.
4. The values of $RRMSEs$ and consequently the values of EFFs for the proposed estimate for different cases indicate that it performs better than the kernel estimate.

5. The proposed estimator $\hat{\Delta}$ performs very well even when the data are simulated from pair Weibull distributions with equal scale or with equal shape parameter. By taking into account that $\hat{\Delta}_k$ is developed without any assumption on the shape of pair distributions, its performance is acceptable but not as that of the proposed one $\hat{\Delta}$. However, we expect that $\hat{\Delta}_k$ may be perform better than $\hat{\Delta}$ if the underlying data distribution is not Weibull.

**Table 1.** The RB, RRMSE and EFF of the estimators $\hat{\Delta}_k$, and $\hat{\Delta}$ when the data are simulated from pair Weibull distributions with equal scale parameters ($\alpha_1 = \alpha_2 = 1$).

| $(n_1, n_2)$ | | $\Delta_{exact}= 0.8678$ $(\beta_1, \beta_2) = (3, 4)$ | | $\Delta_{exact}= 0.6774$ $(\beta_1, \beta_2) = (3, 6.2)$ | |
|---|---|---|---|---|---|
| | | $\hat{\Delta}_k$ | $\hat{\Delta}$ | $\hat{\Delta}_k$ | $\hat{\Delta}$ |
| (20, 30) | RB | -0.2744 | -0.0608 | -0.3009 | 0.0184 |
| | RRMSE | 0.2967 | 0.1271 | 0.3190 | 0.1228 |
| | **EFF** | **1.0000** | **2.3346** | **1.0000** | **2.5976** |
| (50, 50) | RB | -0.2016 | -0.0319 | -0.2018 | -0.0249 |
| | RRMSE | 0.2160 | 0.0753 | 0.2407 | 0.1172 |
| | **EFF** | **1.0000** | **2.8661** | **1.0000** | **2.0543** |
| (100, 200) | RB | -0.1086 | 0.0006 | -0.0560 | 0.0139 |
| | RRMSE | 0.1122 | 0.0503 | 0.0827 | 0.0608 |
| | **EFF** | **1.0000** | **2.2285** | **1.0000** | **1.3587** |
| | | $\Delta_{exact}= 0.4880$ $(\beta_1, \beta_2) = (3, 10.3)$ | | $\Delta_{exact}= 0.2979$ $(\beta_1, \beta_2) = (3, 20.4)$ | |
| (20, 30) | RB | -0.4367 | -0.0795 | -0.5952 | -0.0374 |
| | RRMSE | 0.4609 | 0.1449 | 0.6324 | 0.2506 |
| | **EFF** | **1.0000** | **3.1809** | **1.0000** | **2.5232** |
| (50, 50) | RB | -0.3203 | -0.0366 | -0.5012 | -0.0488 |
| | RRMSE | 0.3418 | 0.1060 | 0.5250 | 0.1264 |
| | **EFF** | **1.0000** | **3.2245** | **1.0000** | **4.1518** |
| (100, 200) | RB | -0.0707 | 0.0138 | -0.1991 | 0.0103 |
| | RRMSE | 0.1084 | 0.0796 | 0.2314 | 0.0824 |
| | **EFF** | **1.0000** | **1.3620** | **1.0000** | **2.8073** |

**Table 2.** The RB, RRMSE and EFF of the estimators $\hat{\Delta}_k$ and $\hat{\Delta}$ when the data are simulated from pair Weibull distributions with equal scale parameters ( $\beta_1 = \beta_2 = 3$ ).

| $(n_1, n_2)$ | | $\Delta_{exact}= 0.8012$<br>$(\alpha_1, \alpha_2) = (1, 1.2)$ | | $\Delta_{exact}= 0.5783$<br>$(\alpha_1, \alpha_2) = (1, 1.5)$ | |
|---|---|---|---|---|---|
| | | $\hat{\Delta}_k$ | $\hat{\Delta}$ | $\hat{\Delta}_k$ | $\hat{\Delta}$ |
| ( 20, 30 ) | RB | -0.2269 | -0.0538 | -0.2052 | -0.0894 |
| | RRMSE | 0.2526 | 0.1264 | 0.2494 | 0.1540 |
| | **EFF** | **1.0000** | **1.9990** | **1.0000** | **1.6192** |
| ( 50, 50 ) | RB | -0.1382 | -0.0449 | -0.0874 | -0.0301 |
| | RRMSE | 0.1622 | 0.1192 | 0.1543 | 0.1156 |
| | **EFF** | **1.0000** | **1.3610** | **1.0000** | **1.3355** |
| ( 100,200 ) | RB | -0.0723 | -0.0109 | -0.0420 | -0.0201 |
| | RRMSE | 0.0893 | 0.0526 | 0.0956 | 0.0815 |
| | **EFF** | **1.0000** | **1.6964** | **1.0000** | **1.1737** |
| | | $\Delta_{exact}= 0.4247$<br>$(\alpha_1, \alpha_2) = (1, 1.8)$ | | $\Delta_{exact}= 0.1071$<br>$(\alpha_1, \alpha_2) = (1, 3.5)$ | |
| ( 20, 30 ) | RB | -0.1293 | -0.0021 | -0.2503 | -0.0477 |
| | RRMSE | 0.2810 | 0.2494 | 0.5722 | 0.5193 |
| | **EFF** | **1.0000** | **1.1266** | **1.0000** | **1.1020** |
| ( 50, 50 ) | RB | -0.0762 | -0.0127 | -0.1459 | 0.0125 |
| | RRMSE | 0.2004 | 0.1571 | 0.4373 | 0.3976 |
| | **EFF** | **1.0000** | **1.2748** | **1.0000** | **1.0998** |
| ( 100,200 ) | RB | -0.0428 | -0.0043 | -0.0989 | -0.0461 |
| | RRMSE | 0.1013 | 0.0875 | 0.3231 | 0.2938 |
| | **EFF** | **1.0000** | **1.1576** | **1.0000** | **1.0997** |

**Table 3.** The RB, RRMSE and EFF of the three estimators $\hat{\Delta}_k$, and $\hat{\Delta}$ when the data are simulated from pair Weibull distributions with different scale and different shape parameters.

| ($n_1, n_2$) | | $\Delta_{exact}= 0.8672$ $(\alpha_1, \alpha_2) = (1, 1.2)$ $(\beta_1, \beta_2) = (2, 1.8)$ | | $\Delta_{exact}= 0.6243$ $(\alpha_1, \alpha_2) = (1, 1.5)$ $(\beta_1, \beta_2) = (3, 1.9)$ | |
|---|---|---|---|---|---|
| | | $\hat{\Delta}_k$ | $\hat{\Delta}$ | $\hat{\Delta}_k$ | $\hat{\Delta}$ |
| (20, 30) | RB | -0.2387 | -0.1015 | -0.1796 | -0.0220 |
| | RRMSE | 0.2728 | 0.1694 | 0.2498 | 0.1408 |
| | **EFF** | **1.0000** | **1.6103** | **1.0000** | **1.7732** |
| (50, 50) | RB | -0.0925 | -0.0006 | -0.1167 | -0.0353 |
| | RRMSE | 0.1157 | 0.0673 | 0.1497 | 0.1136 |
| | **EFF** | **1.0000** | **1.7187** | **1.0000** | **1.3176** |
| (100,200) | RB | -0.0450 | -0.0120 | -0.0617 | -0.0254 |
| | RRMSE | 0.0644 | 0.0583 | 0.0898 | 0.0658 |
| | **EFF** | **1.0000** | **1.1049** | **1.0000** | **1.3653** |
| | | $\Delta_{exact}= 0.4370$ $(\alpha_1, \alpha_2) = (1, 1.8)$ $(\beta_1, \beta_2) = (4, 2.1)$ | | $\Delta_{exact}= 1646$ $(\alpha_1, \alpha_2) = (1, 3)$ $(\beta_1, \beta_2) = (6, 2)$ | |
| (20, 30) | RB | -0.2465 | -0.0254 | -0.3618 | 0.0263 |
| | RRMSE | 0.3257 | 0.1981 | 0.4706 | 0.3359 |
| | **EFF** | **1.0000** | **1.6443** | **1.0000** | **1.4007** |
| (50, 50) | RB | -0.1510 | -0.0271 | -0.2848 | 0.0067 |
| | RRMSE | 0.2250 | 0.1668 | 0.3588 | 0.2153 |
| | **EFF** | **1.0000** | **1.3488** | **1.0000** | **1.6659** |
| (100,200) | RB | -0.0779 | -0.0064 | -0.2040 | -0.0339 |
| | RRMSE | 0.1234 | 0.0802 | 0.2339 | 0.1239 |
| | **EFF** | **1.0000** | **1.5384** | **1.0000** | **1.8881** |

# References

Alodat, T, Al Fayez, M. and Eidous, O. (2021). On the asymptotic distribution of Matusita's overlapping measure. *Communications in Statistics - Theory and Methods*, 51 (20), 6963-6977.

Al-Saidy O, Samawi H.M., Al-Saleh M.F. (2005). Inference on overlapping coefficients under the Weibull distribution: Equal shape parameter. *ESAIM: Probability and Statistics*, 9, 206-219.

Arif, O. H., and Eidous, O. (2017). Fourth-order kernel method for simple linear degradation model. *Communications in Statistics-Simulation and Computation*, 47 (1), 16-29.

Ba Dakhn, L. N., Al-Haj Ebrahem, M. and Eidous, O. (2017). Semi-parametric method to estimate the time-to-failure distribution and its percentiles for simple linear degradation model. *Journal of Modern Applied Statistical Methods*. 16(2), 322-346.

Chaubey, G., Metspalu, M., Karmin, M. and Thangaraj, K. (2008). Language shift by indigenous population: A model genetic study in South Asia. *International Journal of Human Genetics*, 8(1), 41.

Dhaker, H., Ngom, P. and Mbodj, M. (2019). Overlap coefficients based on Kullback-Leibler divergence: exponential populations case. *International Journal of Applied Mathematical Research,* 6(4), 135-140.

Eidous, O. (2009). Kernel method starting with half-normal detection function for Line transect density estimation. *Communications in Statistics-Theory and Methods*, 38, 2366-2378.

Eidous, O. (2011). Variable location kernel method using line transect sampling. *Environmetrics*, 22, 431-440.

Eidous, O. (2012). A new kernel estimator for abundance using line transect sampling without the shoulder condition. *Journal of the Korean Statistical Society*, 41, 267-275.

Eidous, O. M. and Abu Al-Hayja,a, M. (2023a). Estimation of overlapping measures using numerical approximations: Weibull distributions. *Jordan Journal of Mathematics and Statistics* (JJMS), 16(4), 741 - 761

Eidous, O., and Abu Al-Hayja's, M. (2023b). Numerical integration approximations to estimate the Weitzman overlapping measure: Weibull distributions. *Yugoslav Journal of Operations Research*, 33 (4), 699-712.

Eidous, O. and Abu Al-Hayja'a, M. (2023c). Weibull distributions for estimation of Pianka and Kullback-Leibler overlapping measures. *Journal of Mathematics and Statistics Research*, 5 (1), 165.

Eidous, O., and Al-Daradkeh, S. (2022). Estimation of Matusita overlapping coefficient for pair normal distributions. *Jordan Journal of Mathematics and Statistics* (JJMS), 15(4B), 1137 - 1151.

Eidous, O. M. and Al-Shourman, A. (2023). Estimating the Weitzman overlapping coefficient using integral approximation method in the case of normal distributions. *Applied Mathematics, Modeling and Computer Simulation*, 42, 1011-1020.

Eidous, O. M, and AL-Talafha, S. A. (2020). Kernel method for overlapping coefficients estimation. *Communications in Statistics: Simulation and Computation*, 51(9), 5139–5156.

Eidous, O. M., and Ananbeh, E. A. (2024a). Kernel method for estimating overlapping coefficient using numerical integration methods. *Applied Mathematics and Computation*, 462, 128339. https://doi.org/10.1016/j.amc.2023.128339.

Eidous, O. M., and Ananbeh, E. A. (2024b). Kernel method for estimating Matusita overlapping coefficient using numerical approximations. *Annals of Data Science* https://doi.org/10.1007/s40745-024-00563-y.

Eidous, O. M, Marie, M. and Al-Haj Ibrahim, M. (2010). A comparative study for bandwidth selection in kernel density estimation. *Journal of Modern Applied Statistical Methods*, 9 (1), 263-273.

Federer, W. T., Powers, L. and Payne, M. G. (1963). Studies on statistical procedures applied to chemical genetic data from sugar beets. *Technical Bulletin, Agricultural Experimentation Station*, Colorado State University.

Gastwirth, J. L. (1975). Statistical measures of earnings differentials. *The American Statistician*, 29(1), 32-35.

Helu, A., Samawi, H. and Vogel, R. (2011). Nonparametric overlap coefficient estimation using ranked set sampling. *Journal of Nonparametric Statistics*, 23(2), 385— 397.

Ichikawa, M. (1993). A meaning of the overlap pedarea under probability density curves of stress and strength. *Reliability Engineering and System Safety*, 41(2), 203-204.https://doi.org/10.1016/0951-8320(93)90033-U.

Inman, H.F. and Bradley, E.L., (1989). The overlapping coefficient as a measure of agreement between probability distributions and point estimation of the overlap of two normal densities. *Communication in Statistics- Theory and Methods*, 18, 3851-3874.

Madhuri, S. M., Sherry, G. and Subhash, A. (2001). Estimating overlap of two exponential populations. *Proceedings of the Annual Meeting of the American Statistical Association*, 2 81, 848–851.

Matusita, K. (1955). Decision rules, based on the distance, for problems of fit, two samples, and estimation. *The Annals of Mathematical Statistics*, 26(4), 631-640. https://doi.org/10.1214/aoms/1177728422.

Morisita, M. (1959). Measuring of interspecific association and similarity between communities. *Mem. Fac. Sci. Kyushu Univ. Series E*, 3, 65-80.

Mulekar, M. S., and Mishra, S. N. (1994). Overlap coefficients of two normal densities: equal means case. *Journal of the Japan Statistical Society*, 24(2), 169-180.

Pastore, M. (2018). Overlapping: R package for estimating overlapping in empirical distributions. *Journal of Open Source Software*, 3(32).

Pianka, E. R. (1973). The structure of lizard communities. *Annual Review of Ecology and Systematics*, 4(1), 53-74.

Rinne, H. (2008). The Weibull distribution: A handbook. CRC Press, Taylor & Francis Group, New york.

Samawi, H. M. and Al-Saleh, M. F. (2008). Inference of overlapping coefficients in two exponential populations using ranked set sample. *Communication of Korean of Statistical Society*, 15 (2): 147-159.

Schmid F. and Schmidt A. (2005). Nonparametric estimation of the coefficient of overlapping-theory and empirical application. *Computational Statistics & Data Analysis*, 50 (6): 1583-1596.

Wais, P. (2017). A review of Weibull functions in wind sector. *Renewable and Sustainable Energy Reviews*, 70, 1099-1107.

Wang, D. and Tiana, L. (2017). Parametric methods for confidence interval estimation of overlap coefficients. *Computational Statistics & Data Analysis*, 106, 12-26.

Weitzman, M. S. (1970). Measures of overlap of income distributions of white and Negro families in the United States (Vol. 22). US Bureau of the Census.